# Nanoscale mapping of native defects in $T_d$-$MoTe_2$ and their impact on graphene-based van der Waals heterostructures

Authors: Wojciech Ryś[1], Iaroslav Lutsyk[1], Mindaugas Lukosius[2], Przemysław Przybysz[1,3], Paweł Dąbrowski[1], Maciej Rogala[1], Paweł Krukowski[1], Paweł J. Kowalczyk[1]

*[1] Department of Solid State Physics, Faculty of Physics and Applied Informatics, University of Lodz, Pomorska 149/153, 90-236 Łódź, Poland.*
*2 IHP- Leibniz Institute for High Performance Microelectronics, Im Technologiepark 25, Frankfurt (Oder), Germany*
*3 Zernike Institute for Advanced Materials, University of Groningen, Nijenborgh 3, 9747 AG Groningen, The Netherlands*

## Abstract

We investigate native defects and their impact on the electronic structure of $T_d$-$MoTe_2$ and graphene/$T_d$-$MoTe_2$ van der Waals heterostructures using scanning tunnelling microscopy and spectroscopy. STM reveals a high density of circular, hollow surface defects that we attribute to Te vacancies. STS on pristine $MoTe_2$ shows semimetallic, V-shaped dI/dV spectra with subtle low-energy states. The vacancies suppress the tunnelling conductance below the Fermi level without introducing new sharp defect-related resonances, consistent with a vacancy-induced redistribution of Mo–Te spectral weight rather than the appearance of new electronic states. In the heterostructure, graphene preserves these $MoTe_2$ fingerprints with no noticeable graphene doping or charge transfer. Our results demonstrate that isolated native Te vacancies act as spectroscopically weak perturbations to the low-energy electronic structure of graphene. The graphene overlayer further screens their impact on the electronic structure, making graphene/$T_d$-$MoTe_2$ heterostructures intrinsically tolerant to such interface defects.

## Introduction

Van der Waals heterostructures combining graphene with transition metal dichalcogenides (TMDCs) have emerged as a versatile platform for engineering novel electronic and spintronic functionalities in recent years. The ability to combine the wide variety of TMDC properties with the most extensively studied two-dimensional (2D) material – monolayer graphene – via van der Waals interactions opens up a broad design space for novel heterostructures. This, in turn, allows the creation of systems not found in nature, providing a testbed for new physics, e.g., spintronic phenomena [1,2] and topology-driven transport [3,4]. Among the most studied combinations are heterostructures of graphene with TMDCs, particularly those incorporating group VI materials such as $MoS_2$, $WS_2$, and, more recently, the semimetallic tellurides $WTe_2$ [5–11] and $MoTe_2$ [12–16]. These latter materials have attracted substantial attention due to their unique electronic properties. Their orthorhombic $T_d$ phases host type-II Weyl fermions, associated with unconventional transport properties and non-trivial topology. $MoTe_2$ in particular undergoes a structural phase transition from a monoclinic 1T′ phase to the orthorhombic $T_d$ phase at approximately 240 K [17], further enriching its phase diagram and potential device applications. The combination of these materials with the exceptional carrier mobility and electronic properties of graphene creates a platform for exploring new physics and device concepts.

Despite the potential they hold, such heterostructures are still not fully explored. For the $T_d$-$MoTe_2$ and graphene system, additional studies are required to understand and fully utilise its properties. In particular, the strength of proximity effects, the extent of local electronic structure modifications, and the role of the interface in governing these changes remain open questions. In TMDCs, point defects are primarily associated with vacancies, substitutions, adatoms, and atomic displacements [18–21]. For VI group-based TMDCs such as $WTe_2$ and $MoTe_2$, the most stable point defects [22–25], chalcogen vacancies, introduce gap states, which can compromise material performance [21]. These states originate from localised dangling bonds associated with transition-metal d-orbitals and act as electron donors, modifying the local carrier concentration [19]. For example, in the work of Muechler et al. [26], they destroy topological states and lead to a gap opening in $WTe_2$. In the case of substitutional defects, chalcogen atoms can be replaced by chemically similar species, such as group XVII elements F, Cl, Br, I, or XV-group elements N, P, or As, for n- and p-type doping, respectively [19]. On the other hand, atoms with a larger atomic radius than the host species tend to remain adsorbed on the surface as adatoms instead [19].

However, a critical bottleneck limiting the performance of real devices remains inadequately understood: the role of native structural defects at heterostructure interfaces. A comprehensive understanding of how these defects behave at heterointerfaces is essential for both fundamental science and technology development. The significant defect density can introduce new phenomena: modifications of electronic, optical [16,18,19,24,27,28] or magnetic [19,28] properties; changes in material stability [16]; or drive structural phase transitions [16,27,29,30]. Overall, such effects can either enhance [19,21,27,29] or degrade [19,21,27] device efficiency, depending on the specific defect configuration and operating conditions. Not only are defects in the van der Waals heterostructures poorly explored, but pristine $MoTe_2$ in its 1T' and $T_d$ phases and their native defects also remain sparsely characterised. Additionally, a thorough comparison between the characteristics of the defects in both pristine and hybrid systems is required. Such findings could be crucial for the practical implementation of TMDC- and graphene-based devices, yet this topic remains largely unexplored in graphene/TMDC heterostructures.

In this work, we aim to address the unexplored aspects outlined above, focusing in particular on the role of native structural defects at heterostructure interfaces. Using scanning tunnelling microscopy (STM) and scanning tunnelling spectroscopy (STS) measurements, we present a nanoscale study of the $T_d$ (γ) phase of $MoTe_2$ and its corresponding graphene van der Waals heterostructure. We pay special attention to how the native $MoTe_2$ defects manifest in the heterostructure and how they affect the local electronic structure of the hybrid system. It is, to the best of our knowledge, the first atomic-resolution STM/STS study of how native chalcogen vacancies in a $T_d$-$MoTe_2$ substrate influence the local electronic structure of an overlying graphene layer.

## Results & discussion

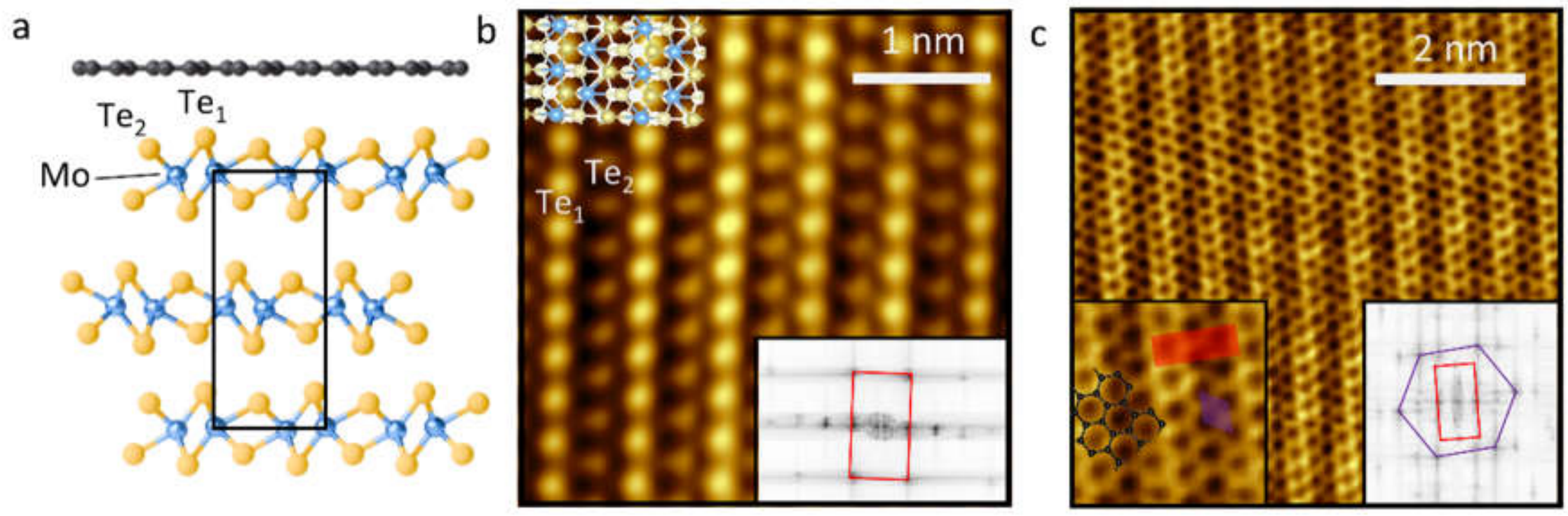

*Figure 1. a) Ball-and-stick side-view model of graphene/$MoTe_2$ heterostructure. Black, blue, and yellow balls correspond to carbon, molybdenum, and tellurium, respectively. b) A high-resolution STM topography of the UHV-exfoliated $T_d$-$MoTe_2$ (I = 0.3 nA; U = 0.15 V). c) The exemplary surface of graphene/$MoTe_2$ (I = 0.2 nA; V = 0.35 V). The FFT and close-up surface images are shown in the insets. The graphene and TMDC surface unit cells are again marked by violet and red shapes, respectively. All data were acquired at ~77 K and FFT-filtered.*

The octahedrally coordinated TMDC $T_d$ phases belong to the orthorhombic crystal system with the $Pmn2_1$ space group (see the ball-and-stick model in Fig. 1a), which underlies the strongly anisotropic in-plane bonding. The experimentally measured lattice constants for this noncentrosymmetric crystal ($T_d$-$MoTe_2$) are: a = 3.477 Å, b = 6.335 Å, c = 13.889 Å [31].

As a reference for the subsequent analysis, we first present nanoscale STM data obtained from bare $MoTe_2$. The pristine $T_d$-$MoTe_2$ surface displays a characteristic striped morphology that originates from alternating rows of upper and lower chalcogen atoms, denoted $Te_1$ and $Te_2$, respectively (Fig. 1b). As shown in Fig. 1b, the clean $MoTe_2$ surface is well resolved, with individual tellurium atoms clearly visible.

Next, we focus on the heterostructure, which consists of monolayer graphene transferred onto an exfoliated molybdenum telluride crystal. Similarly to the bare $MoTe_2$, we again observe the characteristic striped pattern but now mixed with a honeycomb graphene overlayer (see Fig. 1c). Interestingly, overlaying graphene on $MoTe_2$ results in the formation of a superstructure, which is evident in the FFT pattern derived from the image in Fig. 1c. Clearly, a multi-spot pattern is observed, evidencing a twist angle of 8.9° ±2.7° (as measured between the spot corresponding to $\vec{b}$ and the closest graphene one). For such angles, moiré patterns usually manifest only as subtle modulations of atomic intensities [32,33].

The surfaces, like the ones shown above (Fig. 1b-c), act as reference flat areas of the samples that are considered defect-free and provide essential context for understanding how native defects modify surface properties.

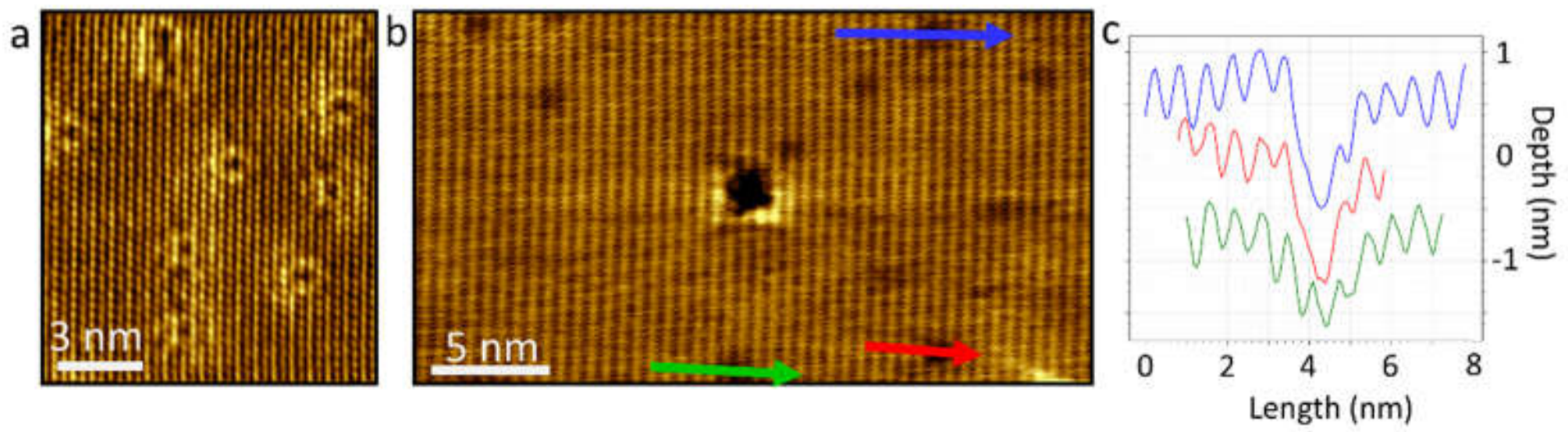

*Figure 2. The images show a set of representative defect topographies for the pristine and hybrid $MoTe_2$ systems. a) Typical pristine surface point-like defects for Td-$MoTe_2$ surface (I = 0.7 nA; U = 0.07 V). b) Round, hollow defects present in the graphene/$T_d$-$MoTe_2$ system (I = 0.05 nA, U = -0.6 V). c) The depth profiles measured along the arrows drawn in "b". All measurements were acquired at ~77 K. All the images shown are FFT-filtered.*

Extended areas that appear free of any defects, such as those shown in Fig. 1, are in fact rare; most regions of the samples contain at least a few defects. Consequently, the most abundant defects in pristine material are hollow circular regions with an optional bright aureola (see Fig. 2a). The density of these defects (based on analysis of several STM images) is estimated to be ~0.04 per $nm^2$ and based on the literature [23,34–39], they are attributed to point-like Te vacancies, which are the most common and low-energy point defects.

In contrast to the bare material, for graphene/$MoTe_2$ we observe dark circular hollow sites, i.e., shallow pits (see Fig. 2b) characterized by a depth reaching ~1 nm (2.5× the atomic modulation, see profiles in Fig. 2c extracted across several defects along lines depicted in Fig. 2b). The density of these defects is of order 0.03 per $nm^2$, i.e., very similar to the bare $MoTe_2$ surface. Therefore, we believe that the origin of these defects is also related to the Te vacancy sites. All observed defects in our experiments were of a similar nature, indicating that the graphene transfer process does not introduce any additional defect types.

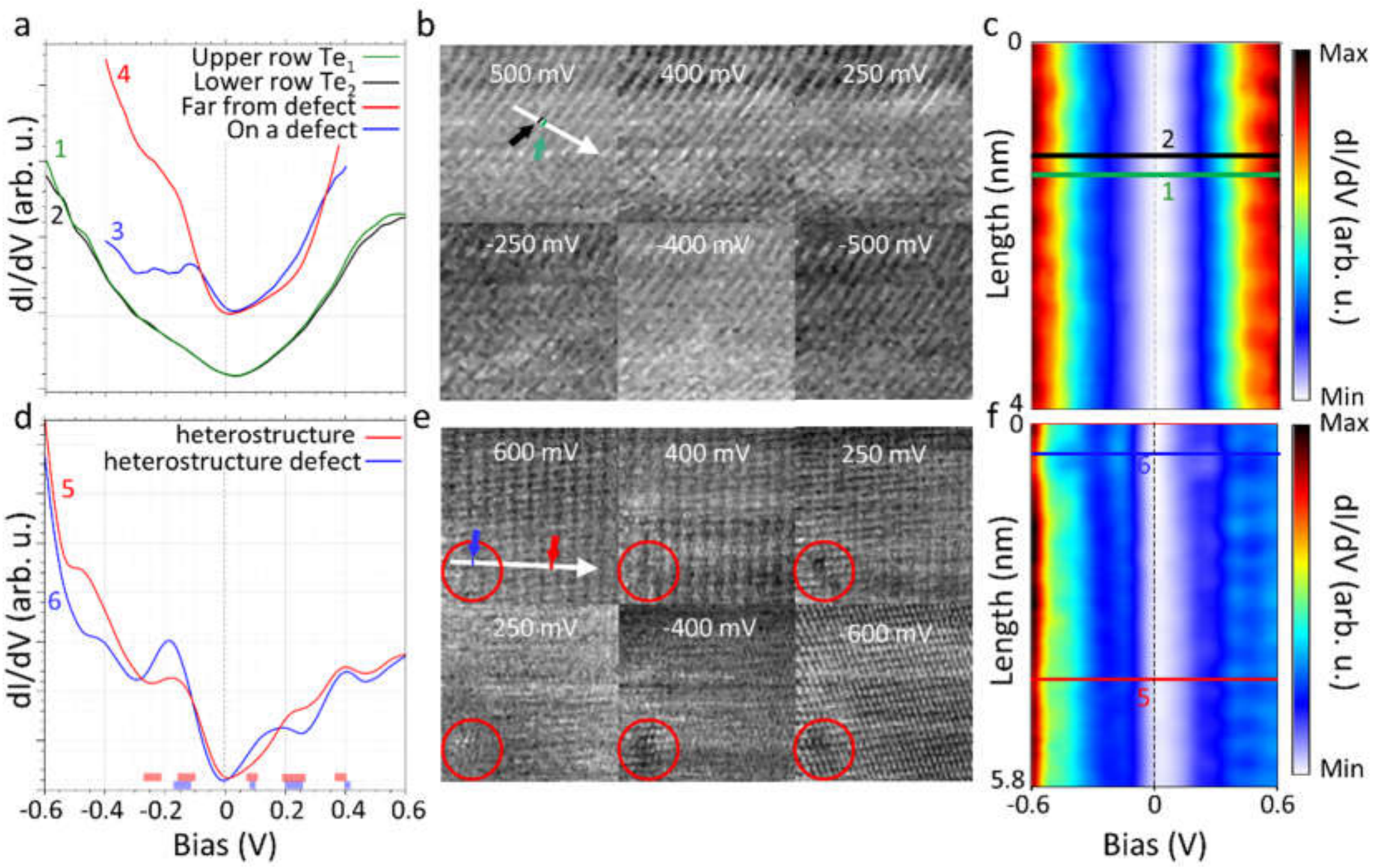

*Figure 3. The STS data sets acquired for the pristine surface found on the $T_d$-$MoTe_2$ (upper row) and for the typical defect found on the graphene/$T_d$-$MoTe_2$ heterostructure (lower row). a) The set of dI/dV curves recorded on bare $MoTe_2$: #1: above $Te_1$ and #2: above $Te_2$ sites (see Fig.1a), #3: far from the defect, #4: in the defect-free region, over a defect. b) A selection of constant-bias dI/dV images (8×8 nm in size) depicting a characteristic striped pattern. c) The dI/dV data averaged along the profile drawn in b. d) The STS curves corresponding to the slices taken along the blue and red cuts of the profile in e-f. e) A selection of constant-bias dI/dV images (7×7 nm in size) depicting patterns characteristic of $T_d$ phases and graphene. f) The dI/dV data averaged along the profile drawn in e. The spectra were acquired at 77 K (I = 0.3 nA; U = 0.5 V and I = 0.08 nA; U = -0.4 V, respectively). Red and blue lines at the bottom of panel d show the location of spectroscopic features, with line lengths corresponding to the standard deviation of maxima read from statistical analysis shown in Fig. S1c (defect-free) and S1d (over defect), respectively. Red circles in "e" mark the defect site.*

In order to better understand defects that are Te vacancies in $MoTe_2$ and graphene/$MoTe_2$, we have used STS to investigate changes in the electronic structure and interfacial interactions. First, we focus on pristine $MoTe_2$ to establish a reference for the hybrid system, particularly given the limited availability of STS data. The UHV-cleaved $MoTe_2$ is characterised by a typical semimetallic V-curve shape (see curves #1-2 in Fig. 3a recorded at $Te_1$ (upper) and $Te_2$ (lower) atomic rows, respectively) with the very subtle $T_d$-$MoTe_2$ states typically located at approximately -0.09 V and 0.13 V. For the statistical distribution of peak maxima locations for all the acquired data, see Fig. S1. The significant statistical uncertainty observed for $MoTe_2$ is mainly caused by frequent tip changes (despite LN conditions), which led to slight variations in the tip's density of states and, consequently, in the measured tunnelling conductance. Additional inhomogeneity between different maps further contributed to this spread.

Moreover, our STS data show a subtle dI/dV intensity modulation related to the $Te_1$ (upper) and $Te_2$ (lower) tellurium rows, clearly seen in a set of constant-bias dI/dV images and a cross-section dI/dV map shown in Fig. 3b and 3c, respectively (see also Fig. S2a for a flattened cross-section dI/dV map). These data show that the modulation is better resolved for positive bias voltages (i.e., above the Fermi level), which is also visible in dI/dV curves extracted from both regions (green and black curves #1-2 in Fig. 3a corresponding to the upper and lower rows, respectively).

Finally, as a reference, we have also recorded a set of spectroscopy data in the region with defects, i.e., above the identified Te vacancies (see curve #3 in Fig. 3a, Fig. S2b and Fig. S3a-c for additional details). The individual dI/dV spectra recorded a few nanometres away from the defect (red curve #4) are similar to the previously discussed ones (green and black, #1-2). In contrast, the dI/dV spectrum recorded on the defect is characterised by a significant decrease in the tunnelling conductance in the negative bias range from approx. -0.1 V downwards (blue curve #3 in Fig. 3a). It thus appears as dark, hollow regions during STM imaging (see Fig. 2a) and in the constant-bias dI/dV images (see Fig. S3a). Within the examined bias range (typically ±0.6 V), no additional states attributable uniquely to the defects are observed, and all the visible features can only be roughly assigned to the same states as in non-defected areas, that are ~ -0.12 V and 0.12 V (see Fig. S1b; note that our statistical analysis points to a noisier character of recorded spectra over defects, which in our opinion is also related to frequent tip changes due to increased tip-sample interaction). Moreover, the global minima of dI/dV spectra remain at the same positions both on and away from the defect. All these results show that the Te vacancy introduces moderate modifications to the electronic structure of $T_d$-$MoTe_2$.

In the next step of our experiments, we have conducted STS measurements on a graphene/$MoTe_2$ hybrid heterostructure. Generally, these measurements reveal more structured spectra, with all states much better resolved than in pristine $MoTe_2$ (see the red curve #5 in Fig.

3d). Taking into consideration some level of variation in the measured STS line shapes, we still find statistically recurring frequent states -0.24 ±0.03 V, -0.13 ±0.03 V, 0.09 ±0.02 V, 0.23 ±0.04 V, and 0.38 ±0.02 V (see Fig. 3d and statistical analysis in Fig. S1c; red lines in Fig. 3d correspond to statistical distribution of maxima i.e., lines length is a standard deviation calculated based on statistical distribution shown in Fig. S1c). These data show some uncertainty in the estimation of maxima number and positions, which reflects the local character of STS measurements.

The striped tunnelling conductance modulation observed in the pristine material is also present in the case of the heterostructure, but only over a limited bias voltage range. Comparing the constant-bias dI/dV images in Fig. 3e with those for pristine $MoTe_2$ (Fig. 3b), the characteristic $MoTe_2$ modulation prevails in the whole positive bias range (see also the cross-section dI/dV map in Fig. 3f and its flattened version in Fig. S2c. The situation changes in the negative-bias regime, where graphene primarily contributes to the constant-bias dI/dV images; the hexagonal lattice is clearly visible, especially for biases smaller than -0.5 V.

Another feature that can be more easily observed in the negative bias range is the defect in the inspected area, which for V < -0.3 V range appears as a dark, roundish region indicated by red circle in Fig. 3e. It is noteworthy that the graphene lattice modulation persists over the defects and appears essentially unaffected (e.g. constant bias dI/dV images recorded for bias equal to -0.4 and -0.6 V in Fig. 3e), hinting that the depression and change in the dI/dV intensity is native to the material below it and not the graphene itself.

Just like for the defects occurring on pristine $T_d$-$MoTe_2$, defects in the heterostructure are characterised by an overall decrease in the tunnelling conductance, mostly for the occupied states below -0.3 V. At the same time, the overall shape of STS curves is maintained, and no new sharp states in the corresponding STS spectra are introduced. The states existing beforehand become shifted and more pronounced, e.g.: -0.14 ±0.03 V (-0.01 shift), 0.09 ±0.01 V (no shift), 0.23 ±0.03 V (no shift) and 0.41 ±0.01 V (+0.03 V shift; compare curve #5 and #6 as well as statistical distribution of maxima represented as blue and red lines in Fig. 3d). This indicates that the Te vacancy exerts a subtle yet non-negligible influence on the local electronic structure of the heterostructure.

The data presented here allow us to draw several general conclusions regarding the influence of Te-vacancy defects on the local electronic structure of $T_d$-$MoTe_2$. In bare $MoTe_2$, the observed spectral changes cannot be described as a simple rigid shift of the electronic structure associated with conventional n-type doping; the observed shifts remain uncertain, lying within a statistical error. A more pronounced effect of the vacancy is the substantial suppression of the local tunnelling conductance for the occupied states below approximately -0.10 V. This reduction may originate from a vacancy-induced modification of the Mo (4d) – Te (5p) hybridization and the accompanying redistribution of spectral weight. It may also partly reflect the removal of the local Te-derived contribution to the LDOS. The reduced spectral background above the vacancy consequently increases the relative contrast of the characteristic $T_d$-$MoTe_2$ spectral features. Their enhanced visibility should therefore not necessarily be interpreted as an increase in their absolute spectral weight, but rather as a result of the suppression of the surrounding continuum of occupied states.

Transferring monolayer graphene onto $T_d$-$MoTe_2$ substantially modifies the tunnelling conditions (as reflected by the change in the statistical distribution in Fig. S1 a-d). The graphene layer increases the effective distance between the STM tip and the underlying $MoTe_2$ surface and consequently reduces the direct contribution of the Mo- and Te-derived orbitals to the tunnelling current. At the same time, the graphene electronic states provide an additional tunnelling channel. As a result, the dI/dV spectra acquired on the heterostructure deviate from the comparatively smooth, V-shaped line shape observed for uncovered $T_d$-$MoTe_2$ and exhibit much more pronounced spectral features. Additionally, performing STS through the graphene overlayer greatly reduces measurement instability, effectively lowering the noise level in the acquired data (compare Fig. S1a–b with S1c–d), likely because the inert carbon layer shields the STM tip from direct physical interaction with the less stable TMDC surface.

Finally, the influence of a Te-vacancy defect located beneath graphene on the measured electronic structure is less pronounced than that observed for a vacancy on uncovered $T_d$-$MoTe_2$. In the heterostructure, the suppression of the tunnelling conductance in the occupied-state region is weaker and becomes apparent mainly below approximately -0.30 V, rather than already below -0.10 V as in bare $MoTe_2$. The principal differences between spectra acquired above and away from the defect are therefore associated with possible minor shifts and changes in the relative prominence of the characteristic spectral features. Importantly, our results suggest that the graphene overlayer does not fundamentally alter the microscopic nature of the defects, although it modifies their spectroscopic manifestation by changing the tunnelling matrix elements, screening the defect potential, and introducing an additional graphene-derived contribution to the tunnelling current.

**Conclusions**

In summary, we have realised and characterised a graphene/$T_d$-$MoTe_2$ van der Waals heterostructure and compared it with pristine $T_d$-$MoTe_2$ at the atomic scale using STM and STS. Our measurements reveal that the dominant native defects on the $T_d$-$MoTe_2$ surface are circular, hollow depressions that we attribute to Te vacancies, with a typical density on the order of $10^{-2}$ per $nm^2$. These same features are observed through the graphene overlayer, demonstrating that native chalcogen vacancies remain structurally and electronically robust at the buried graphene/$MoTe_2$ interface.

STS on pristine $T_d$-$MoTe_2$ shows semimetallic, V-shaped dI/dV spectra with subtle, yet reproducible, states at specific energies. In the presence of Te vacancies, the tunnelling conductance below the Fermi level is reduced, but no additional sharp defect resonances appear within the examined bias window, indicating that these native defects mainly modulate the amplitude of the electronic states rather than generating new states. This behaviour is consistent with a vacancy-induced redistribution of Mo–Te spectral weight and a corresponding suppression of the occupied-state continuum, rather than the emergence of new electronic states. When graphene is added on top of $MoTe_2$, the overall spectra become more structured and show reduced noise, possibly due to increased STM tip stability during measurements. Still, the characteristic low-energy features remain similar to those of pristine $MoTe_2$, indicating low influence of graphene on the electronic structure of $MoTe_2$.

The combined STM/STS paints a coherent picture of the defect–graphene interaction. Notably, the suppression of tunnelling conductance above Te vacancies is markedly weaker beneath graphene than on bare $MoTe_2$, indicating that the graphene overlayer partially screens the defect potential and modifies the tunnelling matrix elements without altering the underlying microscopic nature of the defect. Overall, isolated native Te vacancies in $T_d$-$MoTe_2$ act as spectroscopically weak perturbations to the low-energy electronic structure of an overlying graphene layer: they slightly modulate the DOS and give rise to shallow local depressions in topography, but do not disrupt the characteristic graphene fingerprints in STS.

From a device perspective, our findings imply that graphene/$T_d$-$MoTe_2$ heterostructures are relatively tolerant to isolated chalcogen vacancies in the $MoTe_2$ substrate, with graphene itself providing an additional layer of protection against their electronic impact. Such defects are unlikely to dominate charge-carrier scattering in the graphene channel or to introduce strong mid-gap states at the interface, at least at the densities studied here. Future work could be enhanced by systematically exploring different classes of defects and other TMDC substrates to establish more general design rules for defect-tolerant graphene/TMDC heterostructures and to identify regimes where defect engineering can be used deliberately to tailor interfacial electronic and spintronic properties.

**Methods**

Bulk crystals used for heterostructure fabrication were prepared from 1T'-$MoTe_2$ purchased from HQ Graphene. Germanium-grown graphene was produced by IHP - Leibniz Institute for High Performance Microelectronics. It was chosen due to its well-controlled domain structure and superior quality compared to commonly used copper-grown material. During the sample preparation, we first exfoliated the bulk crystal with adhesive tape under the deionised water surface. Then, we transferred the poly (methyl methacrylate)-coated graphene layer onto the water surface and gently removed the liquid, thus depositing the graphene on the crystal. Next, the sample was immersed in warm acetone to remove PMMA and left to dry. This procedure produces clean heterostructures with minimal residual contamination or organic residue, critical for obtaining atomic-resolution STM images [32]. Before STM studies, the graphene/1T'-$MoTe_2$ heterostructure was cooled down to 77 K for the phase transition to the $T_d$ phase (~240 K [17]) to occur.

The scanning tunnelling microscopy and spectroscopy data were acquired at liquid nitrogen temperature using the Scienta Omicron LT STM system. The typically applied bias ranged over ±1 V and the setpoint current ranged from 0.05 to 2 nA. The STM measurements were performed with either a mechanically cut tip made of 90% Pt-10% Ir alloy wires (Goodfellow) or with an etched tungsten tip (Bruker). The data were processed with Gwyddion 2.59. The STM images were corrected for linear thermal drift using a graphene pattern in reciprocal space as a lattice reference [40] for the heterostructures. The topographic images shown were also denoised via high-frequency FFT filter removal. The additional flattened cross-section dI/dV maps corresponding to all main figures from Fig. S2 were generated using Origin Pro. The process included transposition of the data in the form of matrices and then subsequent normalisation of each column, resulting in a new, flattened map. The statistical analysis shown in Fig.3 and Fig. S1, usually consisting of ~16000 curves per map, was done via algorithmic peak and shoulder finding using Modular Analyse System 2.6.7 software [41,42].

**Supporting information**

Supplementary data related to this article can be found online.

**Data availability**

The data that support the findings of this study are available from the corresponding authors upon reasonable request.

**Acknowledgements**

This work is financed by the National Science Center, Poland, under the projects 2018/31/B/ST3/02450 (W.R., I.L., P.J.K.).

Supplementary information to

# Nanoscale mapping of native defects in $T_d$-$MoTe_2$ and their impact on graphene-based van der Waals heterostructures

Authors: Wojciech Ryś[1], Iaroslav Lutsyk[1], Mindaugas Lukosius[2], Przemysław Przybysz[1,3], Paweł Dąbrowski[1], Maciej Rogala[1], Paweł Krukowski[1], Paweł J. Kowalczyk[1]

*[1] Department of Solid State Physics, Faculty of Physics and Applied Informatics, University of Lodz, Pomorska 149/153, 90-236 Łódź, Poland.*
*2 IHP- Leibniz Institute for High Performance Microelectronics, Im Technologiepark 25, Frankfurt (Oder), Germany*
*3 Zernike Institute for Advanced Materials, University of Groningen, Nijenborgh 3, 9747 AG Groningen, The Netherlands*

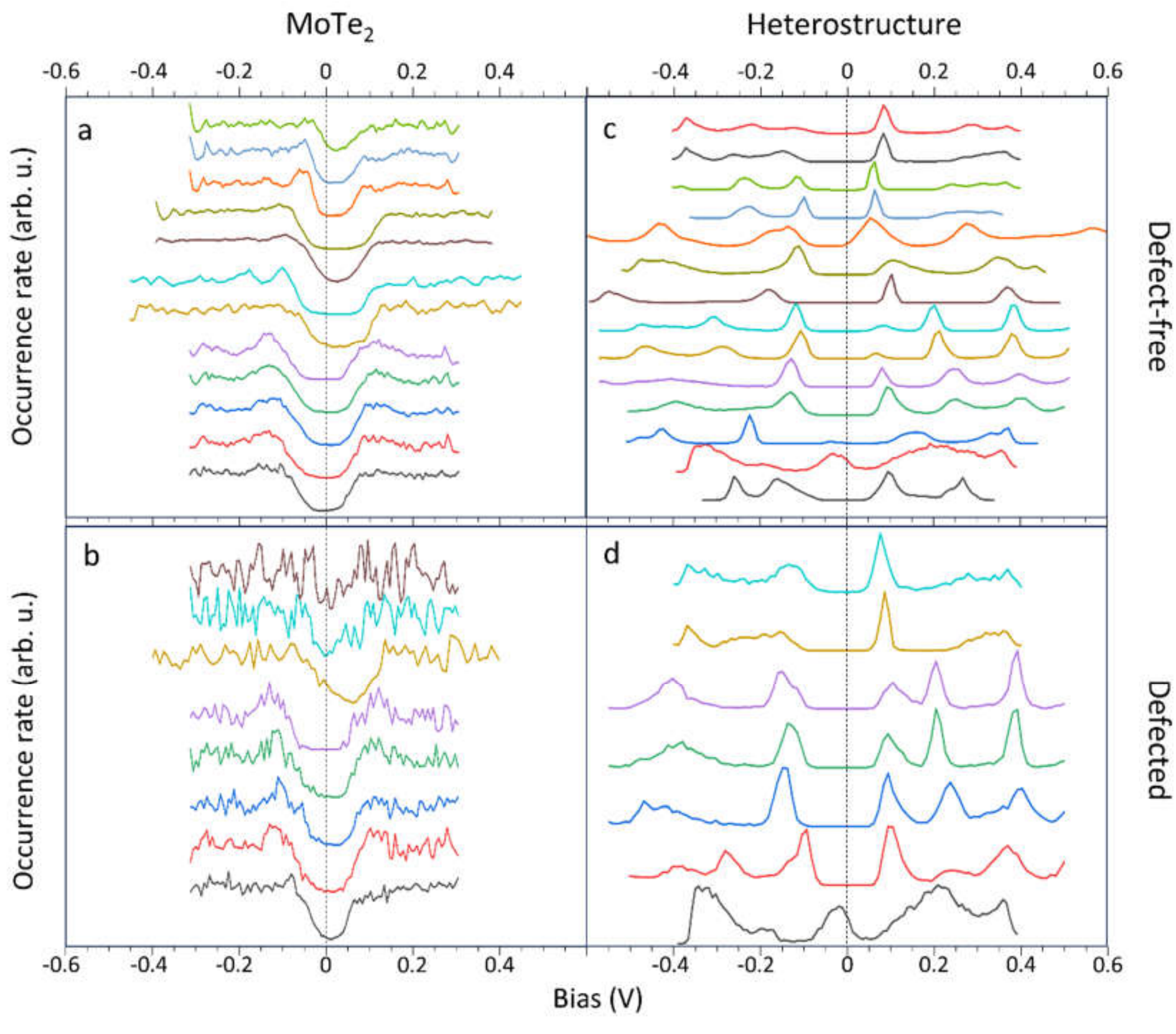

*Figure 1. The set of histograms depicting the occurrence rates of the STS peak maxima for each acquired STS map. a) Data acquired for the pristine $MoTe_2$. b) Data acquired for the $MoTe_2$ defect sites. c) Data acquired for the defect-free surface of the heterostructure. d) Data acquired for the defect sites on the heterostructure. The intensities of the curves in each quadrant are normalised to facilitate comparison.*

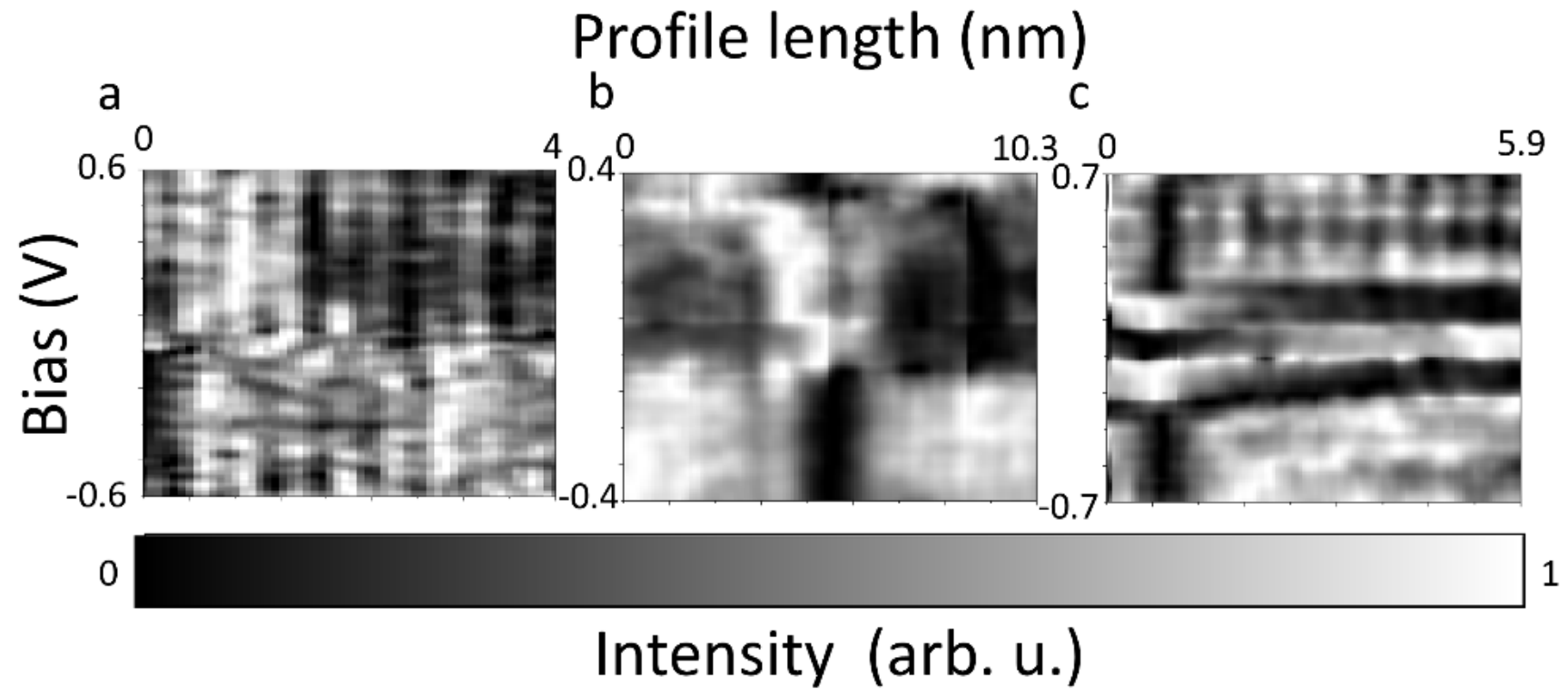

*Figure 2. The flattened STS maps. The spectra originate from all the cross-section dI/dV maps present in the work and in the supplementary information. a) Pristine $T_d$-$MoTe_2$ map from Fig. 3c. b) $T_d$-$MoTe_2$ defect map from Fig. S3b. c) Heterostructure map from Fig. 3f. More details can be found in the methods section of the main text.*

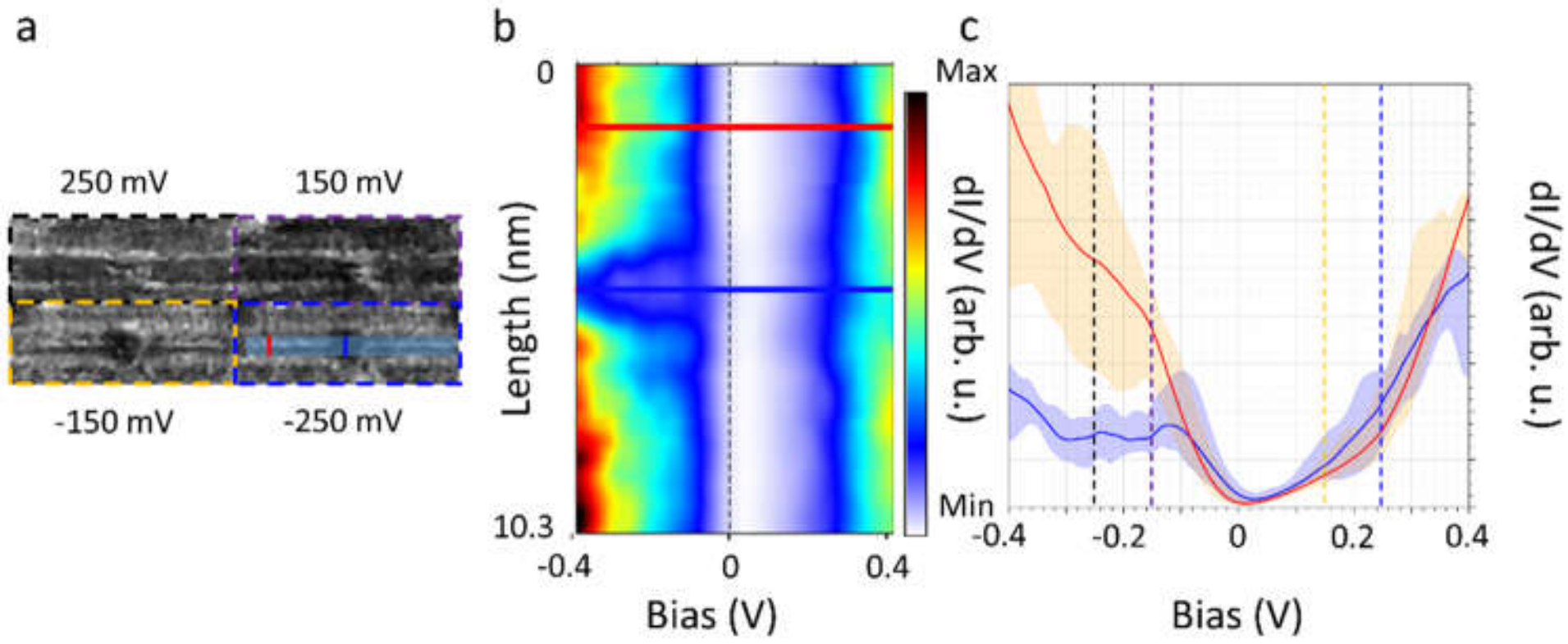

*Figure 3. The STS data sets acquired for a typical defect in $T_d$-$MoTe_2$. a) A selection of dI/dV data slices (5×12 nm in size), b) the dI/dV data averaged along the profile drawn in a and b, c) the STS curves corresponding to the slices taken along the cuts of the profile in a. There is a min/max contour added to the curves, which corresponds to the thinner averaged profiles to highlight the volatility range. The spectra were acquired at ~80 K (I = 0.4 nA; U = -0.08 V).*